\documentclass[aps,reprint, prf, nofootinbib, onecolumn]{revtex4-1}
\usepackage{amsmath}
\usepackage{amssymb}

\usepackage{graphicx}% Include figure files
\usepackage{dcolumn}% Align table columns on decimal point
\usepackage{bm}% bold math
\usepackage{color}%
\usepackage{hyperref}
\newcommand{\la}{\left<}
\newcommand{\ra}{\right>}

\newcommand{\cor}[1]{{#1}}

\newcommand{\nusselt}{\mathrm{Nu}}
\newcommand{\rayleighQ}{\mathrm{Ra_Q}}
\newcommand{\rayleigh}{\mathrm{Ra}}
\newcommand{\prandtl} {\mathrm{Pr}}
\newcommand{\reynolds} {\mathrm{Re}}

\usepackage[normalem]{ulem}
\AtBeginDocument{\let\latexlabel\label}
\newcommand{\paragrephion}[1]{\paragraph*{#1.}}

\begin{document}

%\preprint{}

\title{Convection driven by internal heat sources and sinks: heat transport beyond the mixing-length or ``ultimate'' scaling regime.}% Force line breaks with \\

\author{B. Miquel}
%\email{basile.gallet@cea.fr}
\email{benjamin.miquel@cea.fr}
%\author{Basile Gallet}%
 \affiliation{%
 Service de Physique de l'{\'E}tat Condens{\'e}, CEA, CNRS UMR 3680, Universit{\'e} Paris-Saclay,
CEA Saclay, 91191 Gif-sur-Yvette, France
 }%
 
 \author{S. Lepot}
%\email{basile.gallet@cea.fr}
%\author{Basile Gallet}%
 \affiliation{%
 Service de Physique de l'{\'E}tat Condens{\'e}, CEA, CNRS UMR 3680, Universit{\'e} Paris-Saclay,
CEA Saclay, 91191 Gif-sur-Yvette, France
 }%
 
 \author{V. Bouillaut}
%\email{basile.gallet@cea.fr}
%\author{Basile Gallet}%
 \affiliation{%
 Service de Physique de l'{\'E}tat Condens{\'e}, CEA, CNRS UMR 3680, Universit{\'e} Paris-Saclay,
CEA Saclay, 91191 Gif-sur-Yvette, France
 }%
 
 \author{B. Gallet}
%\email{basile.gallet@cea.fr}
%\author{Basile Gallet}%
 \affiliation{%
 Service de Physique de l'{\'E}tat Condens{\'e}, CEA, CNRS UMR 3680, Universit{\'e} Paris-Saclay,
CEA Saclay, 91191 Gif-sur-Yvette, France
 }%


\begin{abstract}
Thermal convection driven by internal heat sources and sinks was recently shown experimentally to exhibit the mixing-length, or ``ultimate'', scaling-regime: the Nusselt number $\nusselt$ (dimensionless heat flux) increases as the square-root of the Rayleigh-number $\rayleigh$ (dimensionless internal temperature difference). While for standard Rayleigh-B\'enard convection this scaling regime was proven to be a rigorous upper bound on the Nusselt number, we show that this is not so for convection driven by internal sources and sinks. To wit, we introduce an asymptotic expansion to derive steady nonlinear solutions in the limit of large $\rayleighQ$, the Rayleigh-number based on the strength of the heat source. We illustrate this procedure for a simple sinusoidal heat source and show that it achieves heat transport enhancement beyond the mixing-length scaling regime: $\nusselt$ increases linearly with $\rayleigh$ over this branch of solutions. Using rigorous upper bound theory, we prove that the scaling regime $\nusselt \sim \rayleigh$ of the asymptotic solution corresponds to a maximization of the heat flux subject to simple dynamical constraints, up to a dimensionless prefactor. Not only do 2D numerical simulations confirm the analytical solution for the sinusoidal source, but, more surprisingly, they indicate that it is stable and indeed achieved by the system up to the highest $\rayleighQ$ investigated numerically, with a heat transport efficiency orders of magnitude higher than the standard mixing-length estimate. 
\end{abstract}

\pacs{Valid PACS appear here}% PACS, the Physics and Astronomy
\maketitle

Motivated by studies of the Rayleigh-B\'enard (RB) system, a fluid layer enclosed between a hot bottom plate and a cold top one, Malkus put forward the physical idea that turbulent convection may maximize the heat transport subject to simple dynamical constraints (the ``power integrals'')\cite{Malkus}. 
Howard introduced the corresponding variational problem and obtained an upper bound on the heat transport enhancement achieved by RB convection \cite{Howard}. Busse improved the approach by including the incompressibility constraint and deriving the so-called ``multi-$\alpha$'' solutions to the variational problem \cite{Busse}. He also showed that a similar approach can be applied to many transport problems, such as momentum transport in a turbulent shear flow \cite{Busse69,Busse70}. In the 90s, Doering and Constantin introduced an alternate method, the ``background method'', to derive similar bounds for a broad range of turbulent flows \cite{Doering92,Doering94,Doering}. In the context of RB convection, these rigorous upper bounds show that the Nusselt number $\nusselt$ (dimensionless heat flux) increases at most like the square-root of the Rayleigh number $\rayleigh$ (dimensionless temperature difference between the plates) in the large-Rayleigh-number regime \cite{Howard,Busse,Doering}. The scaling behaviour of this upper bound is reminiscent of the ``mixing-length'' arguments put forward in astrophysical contexts, where it is assumed that the relation between the heat flux and the temperature difference does not involve the molecular diffusivities \cite{Spiegel63,Spiegel}. The resulting ``mixing-length'' scaling-law is $\nusselt \sim \sqrt{\prandtl \, \rayleigh}$, where $\prandtl$ is the Prandtl number. In the context of RB convection, this  scaling regime was initially proposed by Kraichnan \cite{Kraichnan}, albeit with additional logarithmic corrections, and is sometimes referred to as the ``ultimate'' regime of thermal convection \cite{Chavanne97}. The actual scaling behaviour $\nusselt \sim \rayleigh^\gamma$ of experimental RB convection remains a controversial issue in the literature, with values of the measured exponent $\gamma$ in the range $\gamma = 0.33-0.39$ for the highest $\rayleigh$ achievable in the laboratory \cite{Chavanne97,Chavanne01,Niemela,Roche,He,Alhers}. This is clearly below the scaling exponent of the upper bound derived by Howard and improved upon by several authors: RB convection does not seem to maximize the heat flux constrained by the power integrals alone. 

It was soon realized that the boundary layers -- either laminar or turbulent -- are responsible for the reduced value of the exponent $\gamma$ \cite{Malkus,Kraichnan,Grossmann}. Various approaches were designed to remove these boundary layers, with the goal of achieving the mixing-length scaling regime predicted by Spiegel for bulk convection inside astrophysical objects \cite{Spiegel63,Spiegel}. The most drastic one consists in numerically simulating ``homogeneous'' convection inside a 3D-periodic domain \cite{Lohse}. While early results of this approach point towards a $\nusselt \sim \rayleigh^{1/2}$ scaling relation, further investigation indicates that the numerical simulations are polluted by diverging elevator-mode solutions at high $Ra$. The latter would grow unboundedly if resolution- and numerical-noise-dependent instabilities did not eventually saturate their amplitude \cite{Calzavarini,DoeringFoF}: at the mathematical level, the consequence is that the Nusselt number can take arbitrarily large values, making the physical relevance of this configuration questionable. \cor{One way around this issue is to consider very tall domains, a situation approached experimentally using convection cells that consist in a narrow vertical channel connecting two reservoirs of hot and cold fluid. Interestingly, the data then displays a square-root relation between the heat flux and the internal temperature gradient \cite{Gibert,Tisserand2010}. However, the associated Rayleigh and Nusselt numbers need be built using the width of the channel\cite{Pawar,Cholemari,Castaing}. The resulting scaling relation between the heat flux and the internal temperature gradient explicitly involves the width of the domain, which challenges its applicability to the arbitrarily wide domains characteristic of astrophysical and geophysical flows. Another line of work consists in replacing the smooth boundaries of RB convection by rough plates. When the roughness has a single scale, it can lead to $\gamma \simeq 0.5$ over a finite range of Rayleigh numbers, the belief being that this range extends to arbitrarily large $\rayleigh$ when the roughness involves infinitely many scales in a fractal fashion \cite{Shen,Ciliberto,Tisserand,Wei,Salort,Goluskin,Xie,Wettlaufer,Zhu,Rusaouen,Zhu2019}.}

%It was soon realized that the boundary layers -- either laminar or turbulent -- are responsible for the reduced value of the exponent $\gamma$ \cite{Malkus,Kraichnan,Grossmann}. Among different approaches designed to remove these boundary layers, the most drastic one consists in numerically simulating ``homogeneous'' convection inside a 3D-periodic domain \cite{Lohse}. While this approach points towards a $\nusselt \sim \rayleigh^{1/2}$ scaling, the corresponding flow {\color{red} exhibits} diverging elevator-mode solutions which saturate due to resolution- and numerical-noise-dependent instabilities~\cite{Calzavarini,DoeringFoF}. {\color{red}Mathematically, the Nusselt number can therefore take arbitrarily large values, rendering this configuration physically questionable.}

Recently, some of the authors developed an alternate convection experiment, where heat is effectively input and extracted in volume through a combination of radiative heating and secular cooling \cite{Lepot,Bouillaut}. This strategy allowed us to bypass the diffusive boundary layers and achieve the mixing-length scaling regime $\nusselt \sim \rayleigh^{1/2}$. The questions that naturally arise are thus: Does convection driven by heat sources and sinks maximize heat transport, and how large is the associated maximum heat flux?
We answer this question by combining two theoretical tools: first, we introduce an asymptotic expansion to derive high-Rayleigh-number steady solutions \cite{Chini,Jimenez,Waleffe}. Second, we derive rigorous upper bounds on the heat transport efficiency of convection driven by internal sources and sinks. Because the scaling behaviour of the asymptotic solution matches the one of the upper bound, we can conclude unambiguously on the maximum heat transport enhancement that can actually be achieved in the system: the asymptotic solution obeys the scaling-law $\nusselt \sim \rayleigh$, which corresponds to heat transport enhancement much beyond the mixing-length scaling regime.

\paragrephion{Convection driven by heat sources and sinks}

Consider a Newtonian fluid of mean density $\rho_0$, specific heat capacity $C$, kinematic viscosity $\nu$ and thermal diffusivity $\kappa$, inside a domain of height $H$. The fluid is subject to internal heat sources and sinks, which enter the heat equation through a term $Q_0 S(z)$, where $Q_0$ characterizes the magnitude of the heat input/output in Watts per cubic meter, $z$ is the vertical coordinate measured in units of $H$, and $S(z)$ is a shape function that depends on the dimensionless vertical coordinate $z$ only. This heating/cooling profile $S(z)$ has zero mean over the fluid domain, to avoid any temperature drift in the long-time limit, and unit root-mean-square (rms) value. After non-dimensionalizing lengths with $H$, time with $H^2/\kappa$, and temperature with $\nu \kappa / \alpha g H^3$, where $\alpha$ is the thermal expansion coefficient and $g$ is gravity, the Navier-Stokes and heat equations in the Boussinesq approximation read:
\begin{eqnarray}
\partial_t {\bf u} + ({\bf u} \boldsymbol{\cdot} {\bm \nabla}) {\bf u} & = & -{\bm \nabla} p + \prandtl ( \Delta {\bf u}+T{\bf e}_z ) \, , \label{equ}\\
\partial_t T + {\bf u} \boldsymbol{\cdot} {\bm \nabla} T & = & \Delta T + \rayleighQ \, S(z) \, . \label{eqT3D}
\end{eqnarray}
In these equations, $t$ and ${\bf x}$ denote the dimensionless time and space variables, while ${\bf u}$, $T$ and $p$ are the dimensionless velocity, temperature, and generalized pressure fields. The velocity field is incompressible, ${\bm \nabla} \boldsymbol{\cdot} {\bf u} = 0$, and the two dimensionless control parameters appearing in the equations are the Prandtl number $\prandtl$ and the flux-based Rayleigh number $\rayleighQ$, defined as:
\begin{eqnarray}
\prandtl=\frac{\nu}{\kappa} \, , \qquad \rayleighQ=\frac{\alpha g Q_0 H^5}{\rho_0 C \kappa^2 \nu} \, .
\end{eqnarray}

%\begin{figure*}[]
%	\includegraphics[scale=0.45]{figures/snapshots_composite.eps}
%	\caption{\label{fig:snapshots} Temperature field in steady state obtained by DNS of equations (\ref{eqPsi}-\ref{eqT}), for no-flux (top row) and vanishing-temperature (bottom row) boundary conditions at top and bottom. The Prandtl number is $\prandtl=1$ and the flux-based Rayleigh number $\rayleighQ$ increases from left to right.}
%\end{figure*}

\begin{figure*}[]
    \includegraphics[width=\textwidth, clip=true, trim={0 0 0 40pt}]{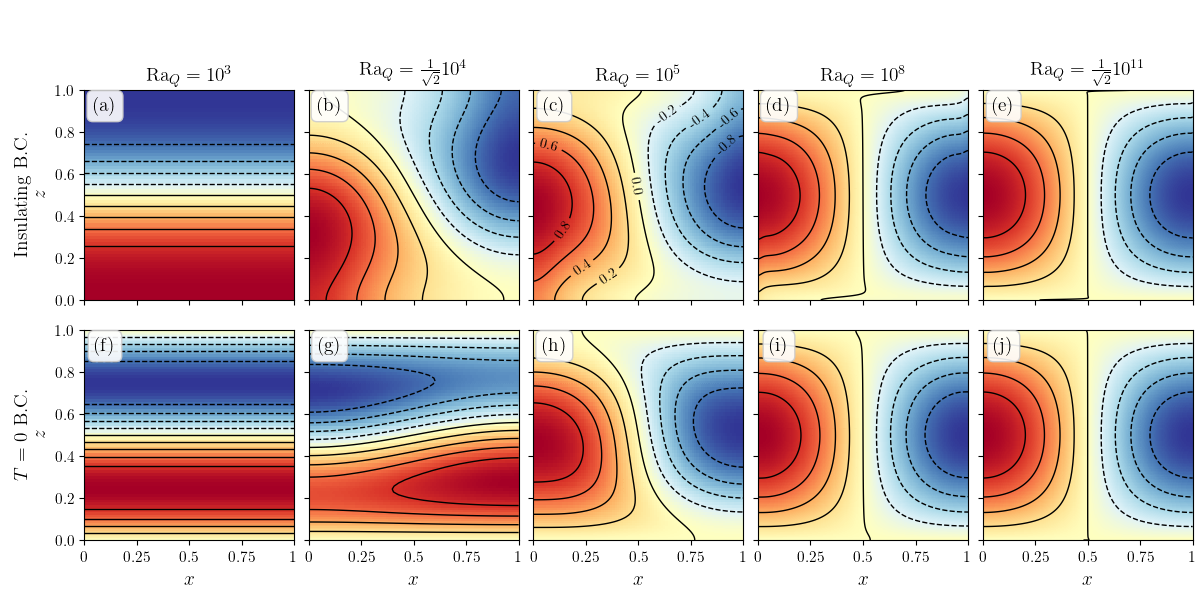}
    \caption{\label{fig:snapshots} Temperature field in steady state obtained by DNS of equations (\ref{eqPsi}-\ref{eqT}), for no-flux (top row [a-e]) and vanishing-temperature (bottom row [f-j]) boundary conditions at top and bottom. The Prandtl number is $\prandtl=1$ and the flux-based Rayleigh number $\rayleighQ$ increases from left to right. All fields are normalized by the spatial maximum $\mathrm{max}\left(T(x,z)\right)$. We represent isocontours for equispaced values indicated in panel (c).}
\end{figure*}

\paragrephion{High-Rayleigh-number asymptotics}

We first consider two-dimensional (2D) asymptotic solutions to equations (\ref{equ}-\ref{eqT3D}) where the fields depend on $x$ and $z$ only. For brevity, we consider a square domain $(x,z)\in\mathcal{D}=[0,1]^2$, although the approach remains valid for an arbitrary aspect ratio. Introducing a streamfunction $\psi$ such that the velocity field is ${\bf u} = {\bm \nabla} \times (\psi \, {\bf e}_y)$, with ${\bf e}_y$ the unit vector along $y$, equations (\ref{equ}-\ref{eqT3D}) yield:
\begin{subequations}\label{eqPsi_and_T}
\begin{eqnarray}
\partial_t \Delta \psi +J(\psi,\Delta \psi) & = & \prandtl \ [\Delta^2  \psi+ \partial_x T] \, , \label{eqPsi}\\
\partial_t T + J(\psi, T) & = & \Delta T + \rayleighQ  \, S(z) \, , \label{eqT}
\end{eqnarray}
\end{subequations}
where the Jacobian is $J(A,B)=\partial_x A \, \partial_z B - \partial_z A \, \partial_x B$. For simplicity, we consider stress-free boundary conditions at all boundaries of the square domain: $\psi=0$, $\Delta \psi = 0$. We consider insulating boundary conditions at the vertical sidewalls ($\partial_x T =0$), and we leave the top and bottom thermal boundary conditions unspecified for now.

We seek a time-independent asymptotic solution at high $\rayleighQ$ and thus expand the fields in powers of $\rayleighQ^{1/2}$:
\begin{eqnarray}
\psi & = & \rayleighQ^{1/2} \psi_0 + \psi_1 + \rayleighQ^{-1/2} \psi_2 + \dots \, ,\\
T & = & \rayleighQ^{1/2} T_0 + T_1 + \rayleighQ^{-1/2} T_2 + \dots \, .
\end{eqnarray}
At order ${\cal O}(\rayleighQ)$, equations (\ref{eqPsi_and_T}) yield respectively:
\begin{subequations}\label{eqPsi_and_T_O1}
\begin{equation*}
\refstepcounter{equation}\latexlabel{eqPsiO1}
\refstepcounter{equation}\latexlabel{eqTO1}
J(\psi_0,\Delta \psi_0) =  0 \, , \qquad J(\psi_0, T_0)  =  S(z) 
\tag{\ref*{eqPsiO1}, \ref*{eqTO1}}
%\label{eqPsi_or_T_O1}\, . \tag{7a,b}
\end{equation*}
\end{subequations}
To lowest order, the flow thus satisfies the Euler equation and advects heat from the sources to the sinks, without the need for diffusive processes. The solution to equation (\ref{eqPsiO1}) is simply 
\begin{eqnarray}
\Delta \psi_0 = F(\psi_0) \, , \label{solO1}
\end{eqnarray}
i.e., the vorticity is a function of the streamfunction, with the only constraint that $F(0)=0$ to satisfy the stress-free boundary conditions. The solutions are very degenerate at this stage: $F$ can be chosen arbitrarily, the corresponding temperature field being then deduced from (\ref{eqTO1}). As we will now see, solvability conditions at the next order constrain the choice of $F$. Equation (\ref{eqPsi}) yields at order ${\cal O}(\rayleighQ^{1/2})$:
%\begin{eqnarray}
%J(\psi_0,\Delta \psi_1) + J(\psi_1,\Delta \psi_0) & = & \prandtl ( \partial_x T_0 + \Delta^2 \psi_0 ) \, . \label{eqPsiO0}
%\end{eqnarray}
%Using (\ref{solO1}), we rewrite this equation as:
\begin{equation}
J(\psi_0,\Delta \psi_1) + J(\psi_1,F(\psi_0)) =  \prandtl ( \partial_x T_0 + \Delta^2 \psi_0 ) \, , \,
\end{equation}
where we have inserted the relation (\ref{solO1}).
Multiplying this equation by $G(\psi_0)$, with $G$ an arbitrary differentiable function, before integrating over the square domain  ${\cal D}$ leads to:
\begin{equation}
 \iint_{\cal D} G(\psi_0) [J(\psi_0,\Delta \psi_1) + J(\psi_1,F(\psi_0)) ]\mathrm{d}x \mathrm{d}z 
 =  \prandtl \iint_{\cal D} G(\psi_0)( \partial_x T_0 + \Delta^2 \psi_0 ) \mathrm{d}x \mathrm{d}z \, . \label{CStemp}
\end{equation}
%For any three functions $a$, $b$ and $c$ that vanish at the boundaries of ${\cal D}$, the functions can be swapped inside triple products of the form $\iint_{\cal D}a J(b,c) \, \mathrm{d}x \mathrm{d}z$ at the expense of changing the sign. 
For any three functions $a$, $b$ and $c$, one of which vanishes at the boundaries of ${\cal D}$, the functions can be swapped inside triple products of the form $\iint_{\cal D}a J(b,c) \, \mathrm{d}x \mathrm{d}z$ at the expense of changing the sign. Using this property, one can prove that
%Indeed, we have $aJ(b,c)+bJ(a,c)=J(ab,c)$, whose integral over the domain vanishes if either $a=0$, $b=0$ or $c=0$ on the boundary of ${\cal D}$.
%The integrand of the first term in (\ref{CStemp}) can thus be replaced by $- J(\psi_0,G(\psi_0))\Delta \psi_1 = 0$, and the integrand of the second term by $-\psi_1 J(G(\psi_0),F(\psi_0))=0$. We conclude that
 the left-hand side of (\ref{CStemp}) vanishes, giving the following constraints on $\psi_0$ and $T_0$:
\begin{equation}
\iint_{\cal D} G(\psi_0)( \partial_x T_0 + \Delta^2 \psi_0 ) \mathrm{d}x \mathrm{d}z = 0 \, , \label{CS}
\end{equation}
for any differentiable function $G$.

To summarize, solving this problem consists in finding the function $F$ in (\ref{solO1}), such that $\psi_0$ satisfies the infinite set of solvability conditions (\ref{CS}), where the temperature field $T_0$ is deduced from $\psi_0$ by solving equation (\ref{eqTO1}). This is an intricate task in general, that may not even always admit a solution. In the following, we thus focus on a simple form for the source function $S(z)$ to provide an explicit example of solution to this problem.

\paragrephion{Sinusoidal heating}The procedure described above becomes surprisingly simple in the situation where the source term varies sinusoidally in the vertical direction, with a shape function $S(z)=\sqrt{2} \sin(2\pi z)$. One can look for solutions to equations (\ref{eqPsi_and_T_O1}) that consist of the gravest Fourier mode only:
\begin{subequations}
\begin{eqnarray}
\psi_0 & = & \psi_m \sin(\pi x) \sin(\pi z) \, ,\label{solpsi} \\
T_0 & = & T_m \cos(\pi x) \sin(\pi z) \, . \label{solT}
\end{eqnarray}
\end{subequations}
Substitution into (\ref{eqTO1}) yields  $T_m = 2\sqrt{2}/\pi^2 \psi_m$. At this stage, the amplitude $\psi_m$ remains undetermined, and we have obtained a continuous family of solutions to the undamped equations. The selection of $\psi_m$
	%one particular member of this family -- one value of $\psi_m$ -- 
	occurs through the solvability conditions (\ref{CS}), which are simultaneously ensured if
	%way for all these conditions to be satisfied simultaneously is to ensure
    $\partial_x T_0 + \Delta^2 \psi_0 = 0$. After substitution of (\ref{solpsi}) and (\ref{solT}), this yields $\psi_m=\pm 2^{-1/4} \pi^{-5/2}$ and $T_m=\pm 2^{7/4} \pi^{1/2}$. This completes the determination of the asymptotic solution.

\begin{figure}
	\includegraphics[width=0.5\columnwidth]{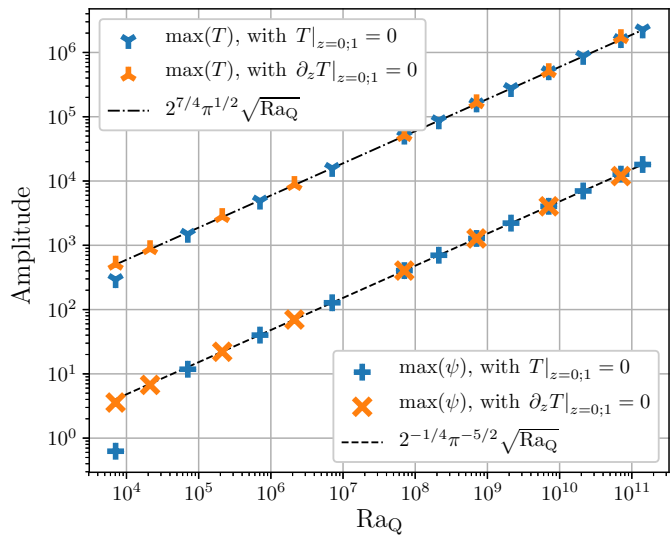}
	\caption{\label{fig:psi_T_vs_RaQ} Spatial maximum of the streamfunction and temperature field as functions of the flux-based Rayleigh number. Symbols correspond to the steady state attained at large time in the DNS with insulating (orange; light grey) or vanishing-temperature (blue; dark grey) top and bottom boundary conditions. The dashed and dash-dotted lines are the predictions $\rayleighQ^{1/2} |\psi_m|$ and $\rayleighQ^{1/2} |T_m|$ from the asymptotic theory.}
\end{figure}

To check the validity of this solution, we performed direct numerical simulations (DNS) of the governing equations (\ref{eqPsi}-\ref{eqT}) with the pseudo-spectral solver \textsc{Coral}~\cite{coral}. We time-stepped the equations inside the domain $[x,z]\in [0,1]^2$ with stress-free boundaries and insulating sidewalls. We consider either vanishing-temperature or insulating boundary conditions at top and bottom. In the latter case, we set the spatially averaged temperature to zero initially, and it remains zero under the evolution equation (\ref{eqT}). %The first surprising fact is that 
Remarkably enough, the system settles in a steady solution in the long-time limit. In figure \ref{fig:snapshots}, we show the temperature field of these steady states for various $\rayleighQ$. The analytical solution derived above readily satisfies $T=0$ at top and bottom, and the numerical solutions indeed tend to the theoretically predicted shape as $\rayleighQ$ becomes larger and larger. For no-flux boundary conditions, small boundary layers develop at the top and bottom to match the analytical solution -- valid in the bulk -- to the no-flux boundary conditions. Methods to compute such boundary-layer corrections in a square box are provided in Refs. \cite{Childress,Chini,Gallet}. In the present situation, these boundary layers have a negligible impact on the temperature field for large $\rayleighQ$, and the solution again takes the shape of the asymptotic analytical prediction, see figure~\ref{fig:snapshots}. For a more quantitative comparison, we plot in figure~\ref{fig:psi_T_vs_RaQ} the spatial maximum of $\psi$ and $T$ in steady state, as functions of $\rayleighQ$. At large $\rayleighQ$, they are in excellent agreement with the predictions $\rayleighQ^{1/2} |\psi_m|$ and $\rayleighQ^{1/2} |T_m|$ of the asymptotic approach.

\paragrephion{Beyond the ultimate regime}

To facilitate the comparison to other convective systems, we now express the asymptotic results in terms of standard Rayleigh and Nusselt numbers. We introduce the Rayleigh number $\rayleigh$ based on the rms temperature, a good proxy for the typical temperature difference inside the square domain. In terms of the dimensionless temperature field, this Rayleigh number is simply $\rayleigh=\sqrt{\la T^2 \ra}$, where $\la \cdot \ra$ denotes space and time average. The Nusselt number is then defined as $\nusselt=\rayleighQ/\rayleigh$.
\cor{Up to a dimensionless prefactor depending on the source function $S(z)$, this number is the ratio of the heat flux transferred from the heat sources to the sinks, to the hypothetical heat flux associated with pure diffusion of the rms temperature over the domain height. Equivalently, $\nusselt$ is proportional to the ratio of the rms temperature of the diffusive steady solution to the rms temperature of the convective one.}

Taking the rms value of (\ref{solT}), remembering that $T = \rayleighQ^{1/2} T_0$ to lowest order, we obtain $\rayleigh=\rayleighQ^{1/2} \, |T_m|/2$. Substituting the expression of $T_m$ and $\rayleighQ=\nusselt \times \rayleigh$ leads to:
\begin{equation}
\nusselt=\frac{\rayleigh}{2^{3/2} \, \pi} \, . \label{scalingNu}
\end{equation}
The asymptotic solution corresponds to heat transport enhancement beyond the mixing-length or ``ultimate'' scaling regime $\nusselt \sim \sqrt{\rayleigh \, \prandtl}$. This new scaling-regime can be traced back to the laminar nature of the flow: the typical velocity is set by the balance between the buoyancy force and the viscous term of the Navier-Stokes equation, which yields a Reynolds number $\reynolds\sim \rayleigh/\prandtl$, as opposed to the standard free-fall estimate $\reynolds \sim (\rayleigh/\prandtl)^{1/2}$ associated to the mixing-length regime of thermal convection \cite{Spiegel63,Spiegel}. This fast flow efficiently carries fluid elements from the heat sources to the sinks, leaving them little time to build large temperature values.

\begin{figure}
	\includegraphics[width=0.5\columnwidth]{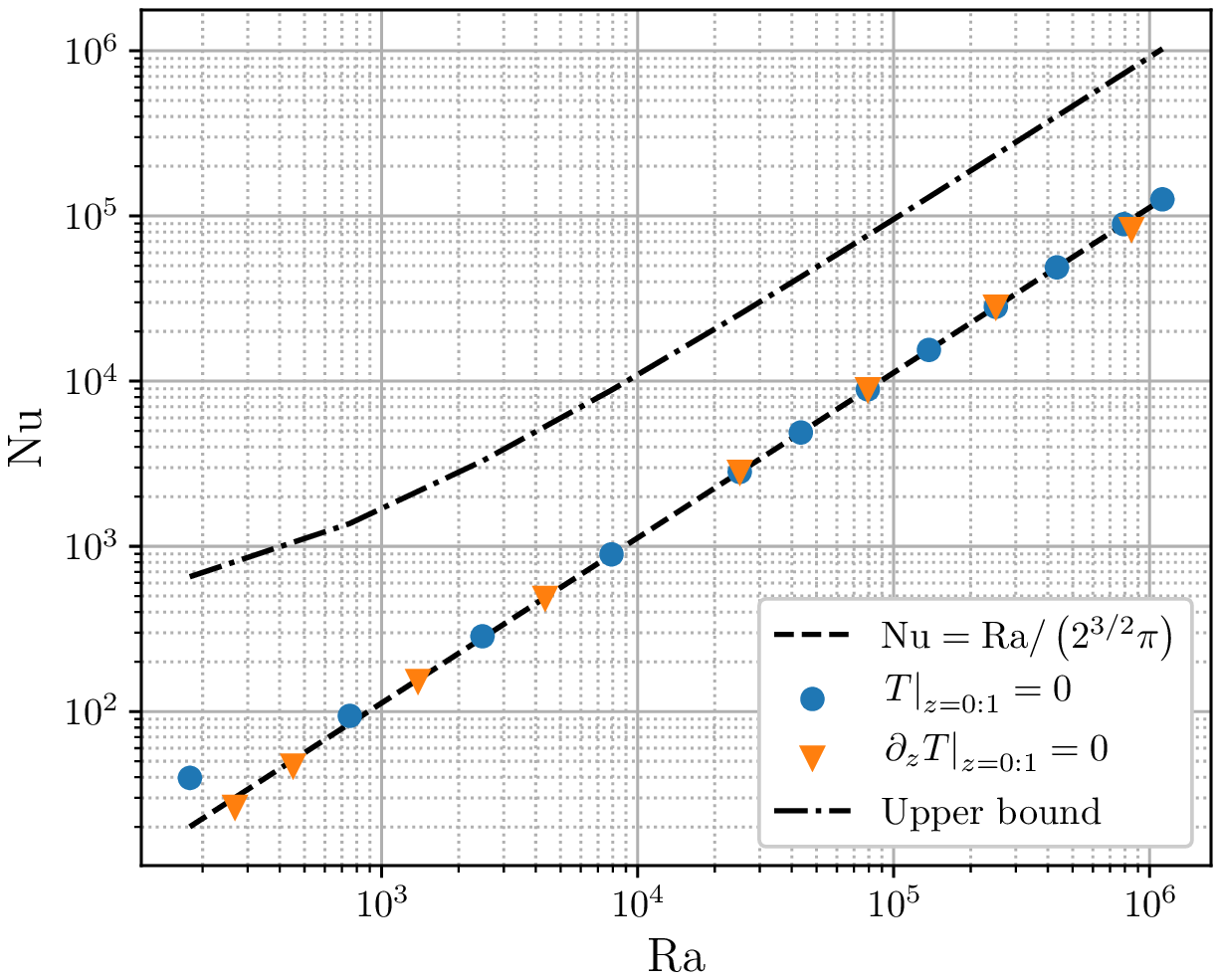}
	\caption{\label{fig:Nu_vs_RaT} Dimensionless heat flux $\nusselt=\rayleighQ/\rayleigh$ as a function of $\rayleigh=\sqrt{\la T^2\ra}$. Same color code as in figure~\ref{fig:psi_T_vs_RaQ}. Dashed line: asymptotic  solution (Eq.[\ref{scalingNu}]). Dash-dotted line: analytic upper bound (Eq.[\ref{bound}]).}
\end{figure}

\paragrephion{Bound on the heat transport}
 
The scaling-law (\ref{scalingNu}) may come as a surprise to readers familiar with Rayleigh-B\'enard convection. Indeed, as mentioned at the outset, for the latter setup one can prove rigorous upper bounds on the Nusselt number of the form $\nusselt \leq {\cal C} \,  \rayleigh^{1/2}$ \cite{Howard,Doering}, where ${\cal C}$ is a dimensionless constant, incompatible with scaling-laws of the form (\ref{scalingNu}). This apparent paradox is solved by observing that %Of course, the simple solution to this apparent paradox is that 
these upper bounds only hold for Rayleigh-B\'enard-like geometries, and not for convection driven solely by internal sources and sinks. One may thus wonder how large the Nusselt number can be for convection driven by heat sources and sinks. We answer this question by deriving a rigorous upper bound on the Nusselt number for this situation. 
We consider the general three-dimensional situation (\ref{equ}-\ref{eqT3D}) inside a parallelepipedic domain of arbitrary aspect ratio, with insulating and either stress-free or no-slip boundary conditions at each boundary. Once again, without loss of generality we set the spatially averaged temperature to zero initially, and it remains zero at any subsequent time.
%with insulating boundary conditions in the horizontal, insulating boundary conditions at top and bottom, and either stress-free or no-slip impenetrable boundary conditions at each boundary (alternatively, one can consider periodic boundary conditions in the horizontal directions). The velocity field is incompressible, so that ${\bm \nabla} \cdot {\bf u} = 0$. We denote as $\la \cdot \ra$ a spatial average inside the domain together with a long-time average. 
Multiplying equation (\ref{eqT3D}) by $S(z)$ before taking a space and time average yields, after a few integrations by parts:
\begin{equation}
\rayleighQ  =  - \la S'(z) w T \ra  - \la S''(z) T\ra  + \left[S'(z) \overline{T}(z)\right]_0^1 \, , \label{tempRaQ}
\end{equation}
where $\overline{T}(z)$ denotes the horizontally- and time-averaged temperature profile, and primes denote $z$-derivatives of the source. We introduce the numerical constants $c_1=\max_{z\in[0;1]} |S'(z)|$ and $c_2=\sqrt{\la S''(z)^2\ra}$ and apply the Cauchy-Schwarz inequality to equation (\ref{tempRaQ}) to obtain:
\begin{equation}
\rayleighQ \leq c_1 \left( \sqrt{\la w^2 \ra} \rayleigh + |\overline{T}(0)| + |\overline{T}(1)| \right) + c_2 \rayleigh \, .
\end{equation}
In a similar fashion, multiplying (\ref{eqT3D}) by $T$ before averaging yields:
\begin{equation}
\la |{\bm \nabla} T|^2 \ra = \rayleighQ \la S(z) T\ra \leq \rayleighQ \, \rayleigh \, ,
\end{equation}
where we have used $\la S(z)^2 \ra =1$ and $\sqrt{\la T^2 \ra}=\rayleigh$.

Because $\la T \ra=0$, there is a $z_0 \in [0;1]$ such that $\overline{T}(z_0)=0$. Hence $|\overline{T}(1)|=|\int_{z_0}^1 \partial_z \overline{T} \mathrm{d}z| \leq \sqrt{\int_{z_0}^1 (\partial_z \overline{T})^2 \mathrm{d}z} \leq \sqrt{\la |{\bm \nabla} T|^2 \ra} \leq \sqrt{\rayleighQ \, \rayleigh}$, and similarly $|\overline{T}(0)|  \leq \sqrt{\rayleighQ \, \rayleigh}$, so that:
\begin{equation}
%\rayleighQ \leq c_1 \left( \sqrt{\la w^2 \ra} \rayleigh +2 \sqrt{\la |{\bm \nabla} T|^2 \ra} \right) + c_2 \rayleigh \, .
\rayleighQ \leq c_1 \left( \sqrt{\la w^2 \ra} \rayleigh +2 \sqrt{\rayleighQ \, \rayleigh} \right) + c_2 \rayleigh \, . \label{ineqRaQ}
\end{equation}
Dotting the Navier-Stokes equation (\ref{equ}) with ${\bf u}$ before averaging \cor{leads to the balance between dissipation and injection of kinetic energy:} $\la |{\bm \nabla} {\bf u} |^2 \ra = \la w T \ra \leq \sqrt{\la w^2 \ra} \rayleigh$. We insert this inequality into the Poincar\'e inequality for $w$, which vanishes at the top and bottom boundaries:
\begin{equation}
\la w^2 \ra \leq \frac{\la |\partial_z w |^2 \ra}{\pi^2} \leq \frac{\la |{\bm \nabla} {\bf u} |^2 \ra}{\pi^2} \leq  \frac{\sqrt{\la w^2 \ra} \rayleigh}{\pi^2} \, .
\end{equation}
Dividing by $\sqrt{\la w^2 \ra}$ we obtain $\sqrt{\la w^2 \ra} \leq \rayleigh / \pi^2$. Inserting the latter inequality into (\ref{ineqRaQ}), substituting $\rayleighQ=\nusselt \times \rayleigh$, finally leads to $\nusselt - c_1 \left( 2 \sqrt{\nusselt} + \rayleigh/\pi^2 \right) -c_2 \leq 0$, which is satisfied only if:
\begin{equation}
\nusselt \leq c_1^2 \left( 1+ \sqrt{1+ \frac{c_2}{c_1^2} + \frac{\rayleigh}{c_1 \pi^2}} \right)^2 \, . \label{bound}
\end{equation}
This inequality is the desired upper bound on the Nusselt number. The right-hand side scales linearly in $\rayleigh$ for large Rayleigh number, which shows that the Nusselt number cannot increase faster than linearly in $\rayleigh$ in the present system. 
For the particular case of sinusoidal heating, the constants entering expression (\ref{bound}) are $c_1=2\sqrt{2}\pi$ and $c_2=4\pi^2$. The upper bound becomes $\nusselt \lesssim 2\sqrt{2}\, \rayleigh /\pi$ at leading order. The heat flux $\nusselt= \rayleigh/2\sqrt{2}\pi$ of the asymptotic solution (\ref{solpsi},\ref{solT}) thus saturates the analytical bound up to a dimensionless prefactor equal to $8$ (see figure~\ref{fig:Nu_vs_RaT}), and the bound can only be marginally improved upon.
\cor{One may want to express the bound (\ref{bound}) using Rayleigh and Nusselt numbers even more closely related to the standard definitions of RB convection. For instance, we  introduce a Rayleigh number based on a characteristic maximum temperature $\text{Ra}_\text{max}=\sqrt{\max_z \{\overline{T^2}(z)\}}$, where the overbar still denotes a horizontal and time average, together with the associated Nusselt number $\text{Nu}_\text{max}=\rayleighQ/\text{Ra}_\text{max}$. An application of H\"older's inequality leads to $\rayleigh \leq \text{Ra}_\text{max}$ and $\text{Nu}_\text{max} \leq \text{Nu}$, so that (\ref{bound}) holds when $\rayleigh$ and $\nusselt$ are replaced by $\text{Ra}_\text{max}$ and $\text{Nu}_\text{max}$.}

\paragrephion{Discussion}
We have derived upper bounds on the Nusselt number, as well as steady asymptotic solutions, that exhibit a heat transport scaling-law $\nusselt \sim \rayleigh$, beyond the mixing-length or ``ultimate'' scaling regime. A natural next step would  be to study the stability of such solutions. For the sinusoidal heat source in 2D, the energy of the convective flow concentrates in the gravest Fourier modes. This observation suggests -- and one may be able to prove analytically --  that the asymptotic solution is indeed stable at arbitrarily large $\rayleighQ$, in a similar fashion to 2D flows forced at the largest scale~\cite{Marchioro}. In our numerical computations we reached $\rayleighQ\approx 10^{11}$ and $\reynolds\approx 4.10^4$ without observing signs that the large scale steady solution would become unstable. However, for more general source functions and in the general 3D situation with no-slip boundary conditions, we observe unsteady and turbulent flows in both experiments and DNS, with an average heat flux obeying the mixing-length scaling-law~\cite{Lepot, Bouillaut}. Steady solutions of the form (\ref{eqTO1},\ref{solO1},\ref{CS}) remain useful in this context, because they show that the upper bound (\ref{bound}) is sharp, i.e., it captures the scaling behaviour of some true solutions to the equations. Improving the scaling behaviour of such bounds would thus require a technology to exclude some unstable solutions during the bounding procedure, a formidable task in general \cite{Fantuzzi}. Instead, \cor{we speculate that }the combination of the asymptotic solutions described above with the upper bound (\ref{bound}) \cor{could} be indicative of the scaling behaviour of extreme events of heat transport: keeping in mind the picture of a chaotic system that wanders between unstable solutions and periodic orbits in phase space \cite{Waleffe2001,Kawahara,Wedin}, unstable solutions of the form (\ref{eqTO1},\ref{solO1},\ref{CS}) may set the scaling behaviour for such extreme events, with $\nusselt \sim \rayleigh$ over the duration of a single event. 
%These issues will be the topic of future studies.
\begin{acknowledgments} We thank C.R. Doering, G. Fantuzzi and D. Goluskin for insightful discussions. This work is supported by the European Research Council under grant agreement 757239, and by Agence Nationale de la Recherche under grant ANR-10-LABX-0039.
	\end{acknowledgments}

\end{document}